\documentclass[twocolumn,showpacs,preprintnumbers,amsmath,amssymb]{revtex4}
\usepackage{graphicx}% Include figure files
\usepackage{dcolumn}% Align table columns on decimal point
\usepackage{bm}% bold math

\begin{document}

\preprint{APS/123-QED}

\title{Quantum gas-liquid condensation in an attractive Bose gas}% Force line breaks with \\

\author{Shun-ichiro Koh}
 
 \email{koh@kochi-u.ac.jp}

\affiliation{ Physics Division, Faculty of Education, Kochi University  \\
        Akebono-cho, 2-5-1, Kochi, 780, Japan 
}%

\date{\today}% It is always \today, today,

\begin{abstract}
Gas-liquid condensation (GLC) in an attractive Bose gas
is studied on the basis of statistical mechanics. Using some results in 
combinatorial mathematics, the following are derived: 
(1) With decreasing temperature, the Bose-statistical coherence grows in 
the many-body wave function, which gives rise to  
the divergence of the grand partition function  prior to Bose-Einstein 
condensation. It is a quantum-mechanical analogue to the GLC in a classical gas 
({\it quantum GLC\/}). 
(2) This GLC is  triggered by the bosons with zero momentum. 
 Compared with the  classical GLC, an incomparably weaker attractive 
 force creates it.  For the system showing the {\it quantum GLC\/}, we 
 discuss a cold helium 4 gas at sufficiently low pressure.
\end{abstract}

\maketitle

\section{Introduction}

Gas-liquid condensation (GLC) is one of the universal phenomena occurring both in 
the classical and the quantum world. 
 GLC is regarded as a singularity appearing in 
the pressure-volume curve determined by the equation of state:
 \begin{equation}
Ê
\frac{P}{k_BT}=\lim_{V\to\infty}\frac{\ln Z_V(\mu)}{V},
Ê
\end{equation}
\begin{equation}
\frac{\rho}{k_BT}=\lim_{V\to\infty}\frac{\partial}{\partial\mu}
ÊÊÊÊÊÊÊÊÊÊÊÊÊÊÊÊ \left(\frac{\ln Z_V(\mu)}{V}\right).
\end{equation}
The question is whether the grand partition function $Z_V(\mu)$ shows the 
singularity \cite{may}\cite{yan}.
 In ordinary GLC such as vapor condensation to water, the GLC occurs 
 at relatively high temperature compared with the 
quantized energy. Hence, this GLC belongs to the classical phenomenon.

When GLC occurs at low temperature as in  helium gas, however, the situation is 
different. One must seriously take into account the influence of quantum statistics
 on the  GLC.  In this paper, we call such a GLC {\it quantum gas-liquid condensation \/} to 
 emphasize the key role quantum mechanics plays in the phenomenon, and we
 study its peculiar nature.  
 
  A prototype of the {\it quantum GLC \/} is the GLC in an attractive 
  Bose gas at sufficiently low temperature. 
 For the relationship between the GLC and Bose statistics, it has been often discussed 
in connection with Bose-Einstein condensation (BEC) \cite{hua}.  
 The following points, however, should be noted. (1) No BEC gas with an attractive 
 interaction remains stable in thermal equilibrium. (The velocity of sound
propagating through the condensate becomes  imaginary \cite{bog}.)   
(2) Since one of the two phases does not exist in the BEC phase,  
a transition between them is impossible. (3) Rather, the {\it quantum GLC \/} 
 occurs {\it in the normal phase close to the BEC transition point \/}. In this region, the
coherent many-body wave function is composed of many Bose particles, but it does not yet reach a 
macroscopic size.  With decreasing temperature,  the bosons' kinetic energy 
 approaches zero, but they do not experience a repulsive 
force until their distance reaches the hard-core radius.
 When a weak attractive force acts on such bosons, its influence is drastic.  
 A dilute Bose gas will undergo GLC prior to BEC \cite{sto}. 
 Its remarkable difference from  classical GLC is that,
  under the influence of Bose statistics, a negligibly weak 
  interaction, which plays almost no  role in the classical phenomenon, 
  causes drastic changes of the system.

\begin{figure}
\includegraphics [scale=0.3]{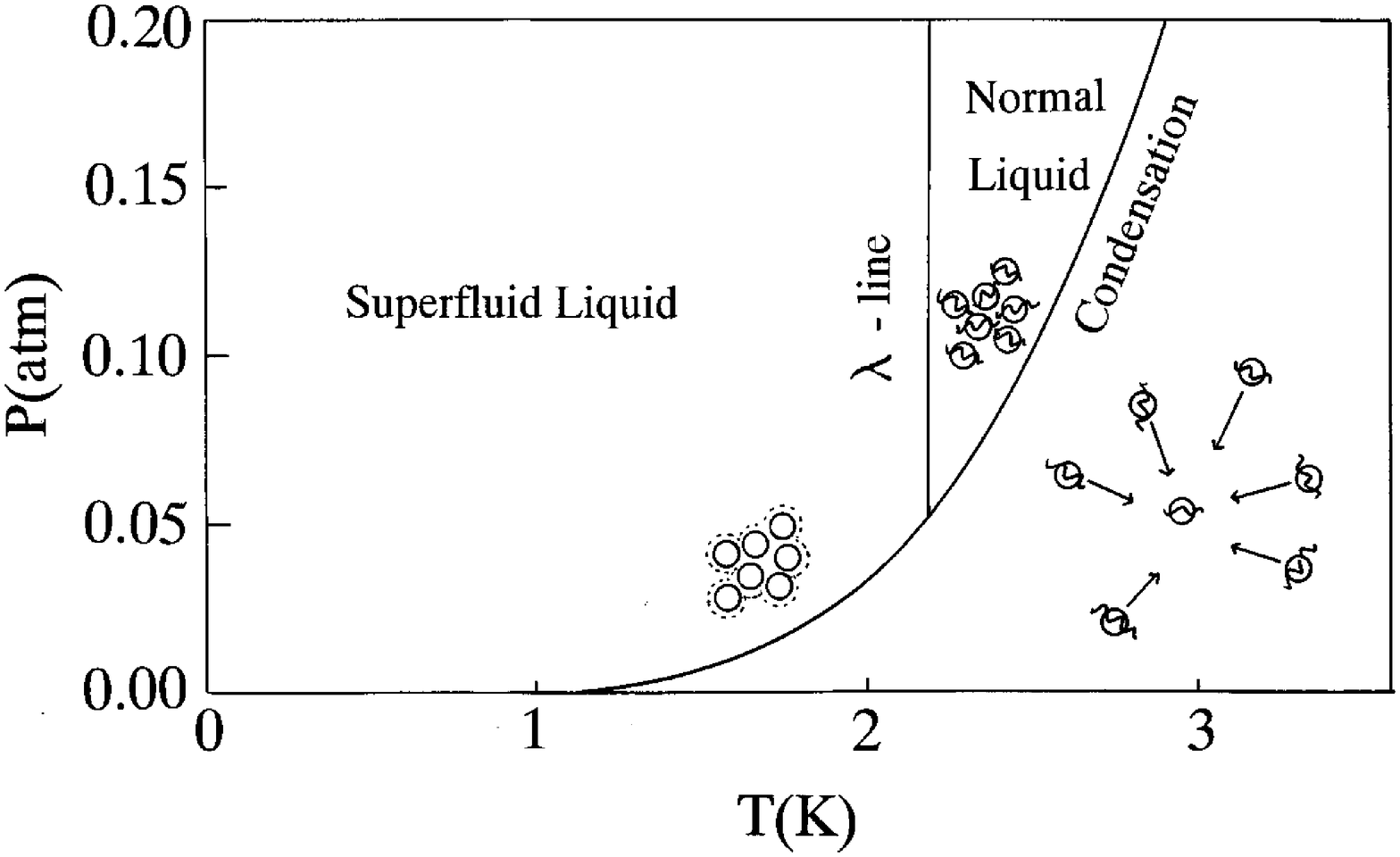}
\caption{\label{fig:epsart}
The phase diagram of helium 4 at low pressure. 
  (In the inset, the wavy lines and the dotted lines around particles
   schematically represent the wave function and the zero-point motion, respectively.) }
\end{figure}

For the system showing the {\it quantum GLC\/}, we know 
an old example. 
 At pressure lower than $0.05$ atm (triple point), a cold helium 4 gas 
  in the normal phase undergoes a phase transition directly to 
   a liquid in the superfluid phase as depicted in Fig.1 \cite{bri}. 
   At sufficiently low pressure, a gas is so rarefied that 
  atoms experience a weak attractive interaction with distant atoms. 
  Because of this weakness, the system remains in the gas state to such a 
  temperature that its kinetic motion is affected by quantum statistics. 
  Hence, it is probable that quantum statistics directly determines the 
  character of this GLC \cite{wel}.

An essential point to formulate the {\it quantum GLC \/} is that 
the coherent many-body wave function, which is symmetrical with respect to interchanges of a large 
number of particles, grows in $Z_V(\mu)$ at low temperature. At the 
vicinity of $T_{BEC}$ in the normal phase, it will give rise to the 
instability of the system. A precise formulation of this quantum 
effect is worth studying closely. 
 The peculiar nature of Bose statistics manifests itself in momentum space. 
Hence, if one formulates the perturbation expansion of $Z_V(\mu)$ in 
momentum space, one may be able to obtain a straightforward expression of 
the instability. Along this line of thought,  statistical mechanics of the {\it quantum GLC \/} was 
 recently considered \cite{koh01}. 

In this paper, using a different method from that in Ref.\cite{koh01} (see Appendix.A), we 
 give a statistical-mechanical model of the {\it quantum GLC \/}. 
  Mayer and others considered the divergence of the perturbation expansion of $Z_V(\mu)$ \cite{may}.
 We will consider the {\it quantum GLC \/} along Mayer's viewpoint \cite{koh03}.
 Using some results in combinatorial mathematics, 
we will derive a set of upper and lower bounds for $Z_V(\mu)$.  
By studying the  radius of convergence for both bounds, 
 we find that the divergence of $Z_V(\mu)$ occurs  prior to BEC in 
 cooling, which leads to the onset of the  {\it quantum GLC \/}. 

  An outline of this paper is as follows. Sec.2 describes the 
 Bose-statistical coherence appearing in $Z_V(\mu)$ at low temperature. 
 Sec.3 develops a perturbation formalism of an attractive 
 Bose gas along the viewpoint of Sec.2, and leads to the instability. 
Sec.4 compares the classical GLC with the {\it quantum GLC \/}, and 
discusses some differences between an attractive Bose and an attractive Fermi gas.

\section{Bose-statistical coherence}
\subsection{A model} 
 
We consider a spinless attractive Bose gas with 
 a weakly attractive  interaction $H_{it}$ 
\begin{equation}
H=\sum_{p}\epsilon _{p}a^{\dagger}_{p}a_{p}
+U\sum_{p,p',q}a^{\dagger}_{p-q}
a^{\dagger}_{p'+q}a_{p'}a_{p},  \quad (U<0) . 
\label{}
\end{equation}
The situation in a low-density gas, in which particles experience 
a negligibly weak interaction since they are far apart, and experience it only when 
a particle encounters other particles, enables us to use a contact 
interaction $U$. To characterize the 
initial stage of the instability occurring in a rarefied gas,
 we ignore the short-range repulsive force which plays an important  
role after the system reaches the high-density state. 
We regard $\sum_{p}\epsilon _{p}a^{\dagger}_{p}a_{p}$ as 
the unperturbed Hamiltonian, and obtain the grand partition 
function $Z_V(\mu)$=Tr $exp[-\beta(H-\mu N)]$ by the perturbation 
theory with respect to the interaction $H_{it}$
\begin{eqnarray}
	  \lefteqn{ Z_V(\mu) = Z_0(\mu) \sum_{n=0}^{\infty}\frac{(-1)^n}{n!} } \nonumber\\ 
        &&\times\int_{0}^{\beta}d\beta_1
  	          \cdots\int_{0}^{\beta}d\beta_n 
	   \langle T H_{it}(\beta _1)\cdots H_{it}(\beta _n)\rangle  ,
\end{eqnarray}¥ 
where  $Z_0(\mu)$ is the grand partition function of an ideal Bose gas.  
($(-1)^n$ comes from the Boltzmann factor.) 
The emergence of the Bose-statistical coherence appears not only in $Z_0(\mu)$ 
but also in $\langle T H_{it}(\beta _1)\cdots H_{it}(\beta _n)\rangle $.

\begin{figure}
\includegraphics [scale=0.4]{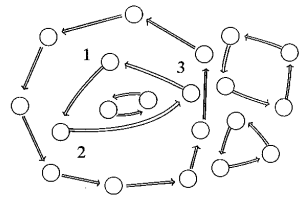}
\caption{\label{fig:epsart}
A pattern of elementary polygons in coordinate space, 
 representing a set of permutations $P$ in the exponent of Eq.(8).
  The size of each polygon reflects a size of the coherent many-body wave 
  function in $Z_0(\mu)$.}
\end{figure}

\subsection{The unperturbed part $Z_0(\mu)$} 
  Figure.2 schematically shows the coherent many-body wave functions in 
coordinate space.  (Thick arrows represent permutation operations of Bose particles.) 
 Matsubara \cite{mat} and Feynman \cite{fey}  associated  the grand partition function
 $Z_0(\mu)=\prod(1-e^{\beta(\epsilon_p-\mu)}¥)^{-1}$ 
with such coherent wave functions as follows.
 The partition function $Z_0(N)$ of $N$ Bose particles is 
given by the density matrix $\rho (x_1,\ldots x_N;x_1',\ldots x_N'; \beta) $ as \cite{tfey}
 \begin{equation}
  	 Z_0(N)=\frac{1}{N!¥} \int\sum_{per}¥
  	       \rho (x_1,\ldots x_N;Px_1,\ldots Px_N; \beta) d^Nx_i¥¥,
  	\label{¥}
  \end{equation}¥  
where $P$ denotes permutations within $N$ particles, and $\rho$ is summed over all possible permutations.
For the unperturbed part,  the density matrix satisfying the ``ideal gas'' equation
\begin{equation}
	\frac{\partial \rho}{\partial \beta¥}
	      =\frac{\hbar^2}{2m¥}\sum_{i=1}^{N¥}\nabla^2_i \rho ¥¥¥
	\label{¥}
\end{equation}¥ 
has  a form as
\begin{eqnarray}
	  \lefteqn{	\rho (x_1,\ldots x_N;x_1',\ldots x_N'; \beta) } \nonumber\\ 
	        &&  =\left(\frac{m}{2\pi\beta\hbar^2¥}\right)¥¥^{3N/2}\exp\left[-\frac{m}{2\beta\hbar^2¥¥}
 	                                                      \sum_{i}^{N¥}¥ (x_i-x'_i)^2¥\right]¥.
\end{eqnarray}¥
One therefore obtains
\begin{eqnarray}
	  \lefteqn{ Z_0(N)=\frac{1}{N!¥}\left(\frac{m}{2\pi\beta\hbar^2¥}\right)^{3N/2} } \nonumber\\ 
	        &&  \times\int\sum_{per}¥  \exp\left[-\frac{m}{2\beta\hbar^2¥}
	            \sum_{i}^{N¥}(x_i-Px_i)^2¥¥¥\right] d^Nx_i¥¥.
\end{eqnarray}¥

 This expression of $Z_0(N)$ is interpreted as follows. 
 For $N=3$ and $P=(1\rightarrow 2, 2\rightarrow 3, 3\rightarrow 1)$, the 
 exponent of Eq.(8) includes $(x_1-x_2)^2+(x_2-x_3)^2+(x_3-x_1)^2$. These 
 permutations represent a triangle in Fig.2, which 
 corresponds to a small coherent wave function. In general, the same
 particle  appears only twice in $\Sigma (x_i-Px_i)^2$ , first at an 
initial position $x_i$ and second at one of the final  positions $Px_i$. 
The intermediate permutations between the initial and final 
$x_s$ form a closed graph, which is visualized as a polygon  
in Fig.2. As a result, one finds that many polygons are spreading out in coordinate 
space. (We call this an {\it elementary polygon\/}.)  The size of  the
polygon  (the number of its sides) corresponds to the number of 
particles, hence the size of the coherent wave function 
obeying Bose statistics  ({\it coherence size\/} \cite{coh}). Each particle belongs to one 
of these polygons, and  polygons do not share the same particle. 
A set of permutations in Eq.(8) corresponds to a pattern of   
polygons like Fig.2. 

Ref.\cite{mat}\cite{fey} calculated $Z_0(N)$ by the following geometrical 
consideration.  A polygon of size $s$ appears in the integrand in Eq.(8) as
\begin{equation}
   ¥\int  \exp\left[-\frac{m}{2\beta\hbar^2¥}(x_{12}^2+\cdots+x_{s1}^2)¥¥¥\right]
             d^{s}x_i¥¥ \equiv f_s.
\label{}
\end{equation}
Assume a polygon distribution as $\{\xi _1,\xi _2, 
\ldots,\xi_s,\ldots \}$, in which elementary polygons of size $s$ 
appear  $\xi_s$ times, being subject to $N=\Sigma _{s}s\xi_s¥$. 
Consider the number of all configurations associated with $\{\xi _s \}$, and denote it with  
 $B(\xi_1,\ldots,\xi_s,\ldots)$. One can rewrite Eq.(8) as 
 
 \begin{eqnarray}
	  \lefteqn{Z_0(N)=\frac{1}{N!¥} \left(\frac{m}{2\pi\beta\hbar^2¥}\right)^{3N/2} } \nonumber\\ 
	        && \times\sum_{\{\xi _s \}}B(\xi_1,\ldots,\xi_s,\ldots)f_1^{\xi_1}\cdots 
                 f_s^{\xi_s}\ldots¥¥.
\end{eqnarray}¥

(1) $B(\xi_1,\ldots,\xi_s,\ldots)$ is estimated as follows \cite{fey}.
Assume $N$ particles in an array.  The number of ways of partitioning them   
 into $\{ \xi _s \}$ is given by $N!/\Pi _s \xi_s!$.  An array of $s$ 
 particles corresponds to a polygon of size $s$. 
For the coherent wave function, it does not matter which particle is an initial one in the 
array (circular permutation). Hence, $N!/\Pi _s \xi_s!$ 
must be multiplied by a factor $1/s$ for each $\xi_s$, with a result that
\begin{equation}
 B(\xi_1,\ldots,\xi_s,\ldots)=\frac{N!}{\prod_s \xi_s! s^{\xi_s}¥}¥.
\label{}
\end{equation}
With Eq.(11) in Eq.(10), one obtains
\begin{equation}
 Z_0(N)=\frac{1}{\lambda^{3N}}¥
        \sum_{\{\xi _s \}}\prod_s\frac{f_s^{\xi_s}}{\xi_s!s^{\xi_s}},
\label{}
\end{equation}
where $\lambda=(2\pi\beta\hbar^2/m)^{1/2}¥$ is the thermal wavelength.

(2) Using Eq.(12), we obtain the grand partition function $Z_0(\mu)=\Sigma _{N}¥Z_0(N)e^{\beta \mu 
N}$, in which summation over $N$ changes to free summation over $\xi_s$ 
from $0$ to $\infty $. Substituting $N=\Sigma _{s}s\xi_s¥$ into 
 $\lambda^{3N}$ in Eq.(12) and $e^{\beta\mu N}$  yields
\begin{equation}
 Z_0(\mu)=\prod_s\exp \left[\frac{f_s}{s}
            \left(\frac{e^{\beta\mu}}{\lambda ^3¥}¥\right)^s¥\right]¥¥.
\label{}
\end{equation}
The $f_s$ in Eq.(9) is estimated with the convolution theorem (see 
Appendix.B), with which one obtains 
\begin{equation}
	 Z_0(\mu) =\exp \left[\sum_{s=1}^{\infty ¥}
	             \left(\frac{e^{\beta\mu s}}{s¥}
	                 +A_s\frac{V}{\lambda ^3}\frac{e^{\beta\mu s}}{s^{5/2}¥}\right)¥\right]¥.
	\label{¥}
\end{equation}¥
The first and the second term in the exponent come from $p=0$ and 
 $p\ne 0$ bosons, respectively, both of which are expansions with respect to the 
coherence size $s$ \cite{as}.
 Using
\begin{equation}
     -\sum_{s=1}^{\infty}\frac{x^s}{s¥}=\ln(1-x) ,	
	\label{¥}
\end{equation}¥
and
\begin{equation}
     \sum_{s=1}^{\infty}\frac{x^s}{s^{5/2}¥}
         =-\frac{4}{\sqrt{\pi }}\int_{0}^{\infty¥}y^2\ln(1-xe^{-y^2})dy¥ ,	
	\label{¥}
\end{equation}
Eq.(14) agrees with $Z_0(\mu)=\prod(1-e^{\beta(\epsilon_p-\mu)}¥)^{-1}$.

Using Eq.(14) in Eq.(1), the coherent many-body wave function appears in 
the equation of states as follows: 
\begin{equation}
\frac{P}{k_BT¥}=\lim_{V\to\infty}\frac{1}{V¥}\sum_{s}¥\frac{ h(s)}{s¥},
\end{equation}
where  $h(s)$ is the size distribution of the coherent wave function 
($1/s$ on the right-hand side is a geometrical factor.). In view of Eq.(14), 
one has $h(s)=e^{\beta\mu s}$ for $p=0$ and $h(s)=(A_sV/\lambda^3) (e^{\beta\mu 
s}/s^{1.5})$ for $p\ne 0$.  The chemical potential $\mu$ determines the 
size distribution. At high temperature ($\mu \ll 0$), $h(s)$ falls off exponentially with a size $s$, 
and large polygons are therefore negligible (Boltzmann statistics). 
 With decreasing temperature ($\mu \rightarrow 0$),
 the exponential dependence on $s$ weakens as $e^{\beta\mu s} \rightarrow 1$, 
 thus making large polygons significant. (At the 
 BEC transition point, its size dependence disappears for $p=0$, and 
 the macroscopic wave function therefore contributes to $Z_0(\mu)$ as much as the 
 microscopic one.)
 It should be noted that, {\it  even in the normal phase, the large but not 
 macroscopic coherent many-body wave functions exist at the 
 vicinity of the BEC transition point\/}, and make considerable contributions to the 
grand partition function.

\subsection{$\langle T H_{it}(\beta _1)\cdots H_{it}(\beta _n)\rangle $}  
In an attractive Bose gas, we must extend the concept of Bose-statistical coherence from 
$Z_0(\mu)$ to $\langle T H_{it}(\beta _1)\cdots H_{it}(\beta _n)\rangle $ in Eq.(4). 

At high temperature,   relatively simple diagrams are significant in 
 $\langle T H_{it}(\beta _1)\cdots H_{it}(\beta _n)\rangle $.  
 A typical example in momentum space is the sum of ring diagrams composed of 
 bubbles  like Fig.3. A thin solid line with an
arrow represents a boson propagator, and dotted lines denote the attractive interaction.

Assume that  $\langle T H_{it}(\beta _1)\cdots H_{it}(\beta _n)\rangle $ is decomposed 
into $h_1+h_2+\cdots $ disconnected diagrams containing closed loops, in which 
each of $h_1$ diagrams contains $m_1$ 
interaction lines, each of $h_2$ diagrams contains $m_2$ interaction lines, and so on 
($n=\sum_{i}h_im_i¥$).  The total number of such diagrams is $n!/h_1!h_2!\cdots$. 
A disconnected diagram with $m$ interaction lines contributes to $Z_V(\mu)$ as
\begin{eqnarray}
	  \lefteqn{\Xi _m=\frac{(-1)^m}{m!} } \nonumber\\ 
        && \times\int_{0}^{\beta}d\beta_1
  	          \cdots\int_{0}^{\beta}d\beta_m 
	   \langle T H_{it}(\beta _1)\cdots H_{it}(\beta _m)\rangle .
\end{eqnarray}¥ 
Hence, a sum of all disconnected diagrams
\begin{equation}
     n!\sum_{h_1}\frac{1}{h_1!¥}\Xi_1^{h_1}\sum_{h_2}\frac{1}{h_2!¥}\Xi_2^{h_2}\cdots 
\end{equation}
replaces the multiple integrals with $(-1)^n$ in Eq.(4), and free sums 
over $h_i$ replace the sum over $n$ as
\begin{equation}
     Z_V(\mu)=Z_0(\mu)\exp\left(\Xi_1+\Xi_2+\cdots\right)¥ .
\end{equation}

\begin{figure}
\includegraphics [scale=0.6]{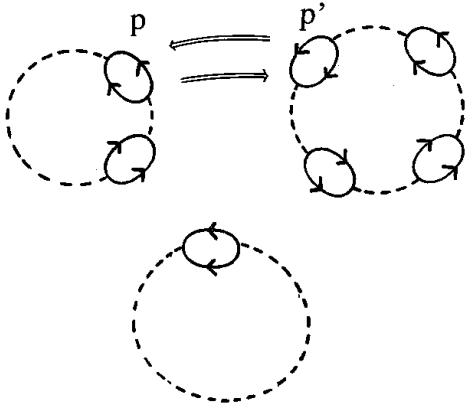}
\caption{\label{fig:epsart}
Some ring diagrams in momentum space, composed of bubbles 
of bosons (an example of $n$=7 terms in Eq.(4)).
 Dotted lines represent  the attractive interaction in Eq.(3). }
\end{figure}

  With decreasing temperature, the wave function in $\langle  T H_{it}(\beta _1)\cdots 
 H_{it}(\beta _n)\rangle $ becomes symmetrical with respect to interchanges of particles. The sum 
 of rings in Fig.3 changes as follows:
When one of two particles in one bubble and that of another bubble  have the same 
momentum ($p=p'$), an exchange of these  
 particles by the thick arrows yields a square as in Fig.4, which links two small rings. 
The inclusion of such diagrams in  $Z_V(\mu)$ ensures the principle of Bose statistics that  
many identical particles have the tendency to occupy the same state ($p=p'$).

\begin{figure}
\includegraphics [scale=0.6]{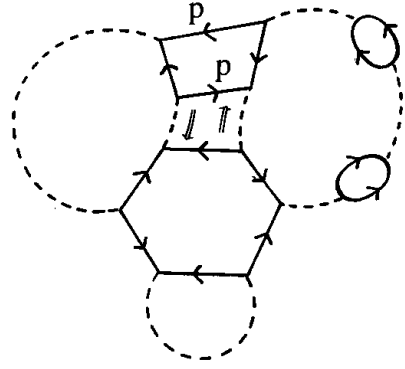}
\caption{\label{fig:epsart}
A polygon cluster including a square and a hexagon, which are 
 made by exchanging particles between two bubbles and between three 
 bubbles in Fig.3, respectively. It belongs to the $(2,1,1,0,0,0,0)$ 
 pattern, an example of $K_1^{\nu  _1} \cdots K_s^{\nu _s}$-type graphs. }
\end{figure}

 Similarly, when three particles are exchanged between three  
 bubbles in the lower part of Fig.3, 
  a hexagon is yielded as in Fig.4. Further, when two particles are interchanged 
between the square and the hexagon in the cluster,  a decagon is yielded.
Such a sequence will continue to a single large polygon in Fig.5. Starting 
from $n$ bubbles, $n$ times of particle interchange between two bubbles 
leads to a single large polygon consisting of $2n$ bosons ($2n$ gon). 
The size of the polygons (the number of its sides) reflects the number of 
particles, hence the {\it size of the coherent wave function\/}.
 With decreasing temperature, the large coherent wave-function becomes important 
 not only in $Z_0(\mu)$ but also in $\langle T H_{it}(\beta _1)\cdots 
 H_{it}(\beta _n)\rangle $. 

In the unperturbed part $Z_0(\mu)$, we consider the elementary polygons. 
Here, we find another type of polygon at each order of the perturbation.  
 From now on we  call this type of polygons  an {\it interaction polygon\/}.
 Both polygons are different, complementary expressions of the coherent many-body wave function.
  In contrast with the elementary polygons, this type of polygon has an 
interaction line at each vertex, thus being connected with each other. 
As depicted in Fig.4, a $2s$-sized polygon is composed of $s$ bosons with $(p_i,l_i)$, 
and another $s$ bosons with $[-(p_i+q_i),-(l_i+m_i)]$, with an expression
given by
\begin{eqnarray}
	  \lefteqn{K_s=\prod_{i=1}^{s¥}U\frac{1}{\beta¥}¥¥} \\ 
	    && \times \left(
	             \frac{1}{\displaystyle {(\epsilon_{p_i}-\mu)+i\frac{\pi l_i}{\beta¥}¥}¥¥}
	             \frac{1}{\displaystyle 
	             {(\epsilon_{-(p_i+q_i)}-\mu)-i\frac{\pi (l_i+m_i)}{\beta¥}¥}¥¥}¥\right)¥¥¥¥, \nonumber	
	\label{¥}
\end{eqnarray}¥ 
where $l,m$ is an even integer including zero.
  While the elementary polygons in $Z_0(\mu)$ represent a 
  series of successive one-particle permutations like Fig.2, the interaction polygons 
represent intertwining of many permutations.

\begin{figure}
\includegraphics [scale=0.6]{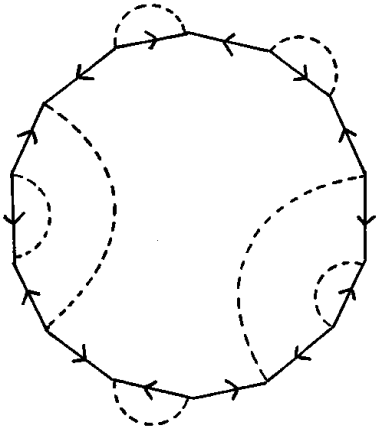}
\caption{\label{fig:epsart}
A single largest polygon (14-gon) as a final result of 
 particle interchanges. It belongs to the $(0,0,0,0,0,0,1)$ pattern, an 
 example of $K_n$-type graphs. }
\end{figure}

\section{Formalism}

To formulate the concept of polygons in $\langle T H_{it}(\beta _1)\cdots 
H_{it}(\beta _n)\rangle $, it is useful to make reference to the description 
below Eq.(9).  In a polygon cluster, let us assume a distribution of interaction polygons 
as $\{\nu _s \} =\{\nu_1,\nu_2, \ldots \}$, in which a polygon $K_s$ of 
size $2s$ appears  $\nu_s$ times.  (For example, $\{  \nu _s \} =\{2,1,1,0, \ldots \}$ in Fig.4.)
  They are connected to each other by $n$ interaction lines, in which 
$2s$ interaction lines emerge from $2s$-sized polygon.
 Hence, $n=\sum_{s}s\nu _s¥$  must be satisfied.

Consider the number of all configurations corresponding to $\{\nu _s \}$, and denote it by 
 $D_n(\nu  _1,\ldots,\nu  _s,\ldots)$.  One can rewrite $\Xi_n$ in 
 Eq.(20) with Eq.(18) in terms of interaction polygons as follows:
\begin{equation}
     \Xi_n=V\frac{1}{n!¥}\sum_{\{\nu _s \}}D_n(\nu  _1,\ldots,\nu  _s,\ldots)
             K_1^{\nu  _1} \cdots K_s^{\nu  _s} \cdots¥¥¥,	
	\label{¥}
\end{equation}¥
which is the analogue of Eq.(10) for $Z_0(N)$.  $H_{it}$ in Eq.(3) is a function of the distance 
 between particles in coordinate space. When the integral in coordinate space is 
performed in $\langle T H_{it}(\beta _1)\cdots H_{it}(\beta _n)\rangle $, 
 relative coordinates of particles are used, 
thus reducing  multiplicity of the integral by one, with a further 
factor $V$ entering in the right-hand side of Eq.(22), which 
represents {\it the translation of the polygon cluster \/}.

For evaluating $\Xi _n$, one must consider the influence of Bose 
statistics on $\prod_{s} (K_s)^{\nu _s}$ and make a graphical consideration of 
$D_n(\nu  _1,\ldots,\nu  _s,\ldots)$.
 
 \subsection{Effects of Bose statistics} 
In $\prod_{s} (K_s)^{\nu _s}$ of Eq.(22),  Bose statistics manifests 
itself in the following manner.

(1) {\it  With decreasing temperature, the feature of Bose statistics, accumulation of particles in the 
 low-energy state,  becomes apparent, thus making the bosons  have a 
 common momentum in each polygon\/}. 
   Hence, one assumes  $p_i=p,l_i=l$, and $q_i=m_i=0$ in Eq.(21), which
 effectively restricts $U$ in Eq.(3) to the pairing-type interaction.
Since each $K_s$ has its own $p$ and $l$ in Eq.(22), 
the summation must be done. One must sum up all cases by replacing each $K_s$ 
in Eq.(22) with $\sum_{l,p}K_s¥$. One therefore redefines $K_s$ in Eq.(22) as
\begin{equation}
	 K_s=\sum_{l,p}
	      \left(-U\frac{1}{\beta¥}¥¥
	             \frac{1}{\displaystyle {(\epsilon_p-\mu)^2+\left(\frac{\pi l}{\beta¥}\right)^2¥}¥¥¥}¥\right)^s
	             \equiv \sum_{l,p}x(p,l)^s¥¥¥. 
	\label{¥}
\end{equation}¥
($(-1)^m$ in Eq.(18) is absorbed in a negative sign in front of $U$ in Eq.(23).)

(2) Let us compare the graph like Fig.4 ($K_1^{\nu  _1} \cdots K_s^{\nu  
_s}$-type) and Fig.5 ($K_n$-type). Since each $K_s$ of $K_1^{\nu  _1} \cdots 
K_s^{\nu  _s}\cdots$ includes a sum over $p$ and $l$ as in Eq.(23),  
$\prod_{s} (K_s)^{\nu _s}$ in Eq.(22) includes the multiple summation.  Hence,  
the number of $K_1^{\nu  _1} \cdots K_s^{\nu  
_s}$-type graphs is larger than that of a $K_n$-type graph.
 With decreasing temperature, however, the feature of Bose statistics, 
 accumulation of particles in the low-energy state, 
 reduces these degrees of freedom.
 {\it  At low temperature, all $K_s$ of $K_1^{\nu  _1} \cdots 
K_s^{\nu  _s}\cdots$  are likely to have a common $p$ and $l$.  \/}
Hence, with decreasing temperature, the various $K_1^{\nu  _1} \cdots K_s^{\nu  
_s}$ terms ($=[\sum x(p_1,l_1)]^{\nu  _1}\cdots[\sum x(p_s,l_s)^{s}]^{\nu 
_s}¥$) approach a single form $K_n=\sum_{p,l}x(p,l)^{n}$, 
because of $n=\sum_{s}s\nu _s¥$. 

\begin{figure}
\includegraphics [scale=0.5]{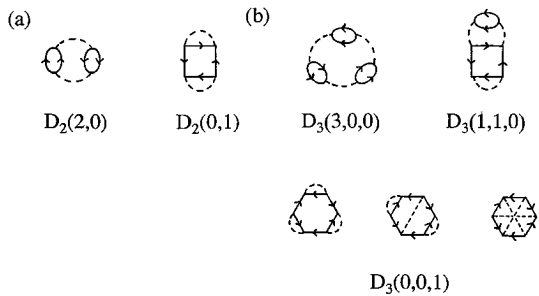}
\caption{\label{fig:epsart}
Polygon clusters. (a) For $n$=2, 
  $D_2(2,0)=1$  and $D_2(0,1)=1$.
  (b) For $n$=3, $D_3(3,0,0)=1$, $D_3(1,1,0)=1$, and $D_3(0,0,1)=3$.}
\end{figure}

 (3) In an attractive Bose gas ($U<0$), $K_s$ is always positive in 
Eq.(23), and $\Xi _n$ in Eq.(22) is therefore  {\it a positive-term series. \/}
 Hence, to derive the upper and the lower bound for $\Xi _n$, one can take only 
 large terms in Eq.(22) without considering the cancellation of terms. 
 (For a repulsive Bose gas, the situation is different \cite{rep}.)

\subsection{Graphical consideration of a polygon cluster}
 For $D_n(\nu  _1,\ldots,\nu  _s,\ldots)$ in Eq.(22),
 it is useful to draw explicitly polygon clusters  in some cases,  
 and get a feeling for its magnitude.  For $n=2,3,4$, we classify the  
 distributions  $\{ \nu _s\}$ satisfying $n=\sum_{s}s\nu _s¥$, 
  and we obtain some $D_n$'s as illustrated in Fig.6 and 7. 
  
\begin{figure}
\includegraphics [scale=0.5]{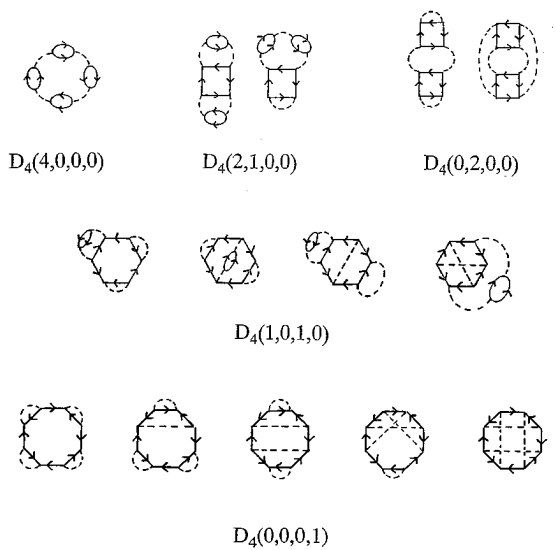}
\caption{\label{fig:epsart}
Polygon clusters for $n$=4: 
 $D_4(4,0,0,0)=1$, $D_4(2,1,0,0)=2$, $D_4(0,2,0,0)=2$, $D_4(1,0,1,0)=4$,  
and $D_4(0,0,0,1)=5$.}
\end{figure}

$n=2$. For $\{ \nu _1,\nu _2 \}$ satisfying 
$2=\nu _1+2\nu _2$, one has $\{ 2,0 \}$ and $\{ 0,1 \}$. 
\\  
 
$n=3$.  For $\{ \nu _1,\nu _2,\nu _3 \}$  satisfying 
$3=\nu _1+2\nu _2+3\nu _3$, one gets $\{ 3,0,0 \}$,$\{ 1,1,0 \}$ and $\{ 0,0,1 \}$.
\\  

$n=4$. For $\{\nu _1,\nu _2,\nu _3,\nu _4\}$
 satisfying  $4=\nu _1+2\nu _2+3\nu _3+4\nu _4$, one obtains $\{ 4,0,0,0 
 \}$, $\{ 2,1,0,0 \}$, $\{ 0,2,0,0 \}$, $\{ 1,0,1,0 \}$, and $\{ 0,0,0,1 \}$. 
\\

 Fig.6(a) shows $D_2(2,0)=1, D_2(0,1)=1$. 
 
 Fig.6(b) shows $D_3(3,0,0)=1$, $D_3(1,1,0)=1$ and $D_3(0,0,1)=3$.

 Fig.7 shows  $D_4(4,0,0,0)=1$, $D_4(2,1,0,0)=2$, $D_4(0,2,0,0)=2$, $D_4(1,0,1,0)=4$,  
and $D_4(0,0,0,1)=5$.

In view of Fig.6 and 7, one notices the following:

(1) In polygon clusters with $n$ interaction lines, there are many 
types of $\{ \nu _s \}$ ranging from $\{n,0,\ldots,0\}$ to $\{0,\ldots,0,1\}$.
 From now on, we call a group of graphs with the same distribution $\{ 
 \nu _s \}$ a ``{\it pattern.\/}'' (For $n$=4, there are five patterns in 
 Fig.7.)  The number of patterns $p(n)$ appearing in the 
$n$ th-order graphs is equivalent to the number of ways of writing $n$ 
 as a sum of positive integers ($n=\sum_{s}s\nu _s¥$).  In  
 combinatorial mathematics, this problem is known as 
  ``partition of $n$'' \cite{han}. It is known 
 to have the following asymptotic form as $n\rightarrow \infty $:
\begin{equation}
       p(n)\sim  \frac{1}{4\sqrt{3}n¥}\exp\left(\pi\sqrt{\frac{2n}{3¥}¥}\right)¥¥.
	\label{¥}
\end{equation}¥

(2) At a given $n$, the pattern with the largest number of 
configurations $D_n$ consists of the largest polygons in size 
(the $2n$-gon,  $\{0,\ldots,0,1\}$).  We can 
confirm that this tendency becomes more evident as $n$ increases. From now on, we abbreviate  
$D_n(0, \ldots,0,1)$ as $D(n)$. (Fig.4 and 5 are
 examples of polygons with seven interaction lines.  Due to various 
arrangements of dotted lines, there are more  Fig.5-type graphs 
than Fig.4-type ones.)

 Combining  $p(n)$ and $D(n)$, one derives an inequality for $\Sigma D_n$ 
 at a given $n$ in $n\rightarrow \infty $,
 \begin{eqnarray}
	  \lefteqn{  D(n) < \sum_{\{\nu _s \}}D_n(\nu  _1,\ldots,\nu  _s,\ldots) } \nonumber\\ 
	        &&   < \frac{1}{4\sqrt{3}n¥}\exp\left(\pi\sqrt{\frac{2n}{3¥}¥}\right)¥D(n),
\end{eqnarray}¥ 
where the lower bound is a case in which the sum over $\{\nu _s \}$ in $\Sigma D_n$ includes 
only a pattern composed of the largest polygons, and the upper bound is a 
case in which all patterns at a given $n$ are included as if their $D_n$ 
is the same as $D(n)$ of the largest polygon.

(3) With decreasing temperature, large polygons, which are successively 
produced by particle interchanges between small polygons, become significant. 
 Because the ``pairing interaction'' becomes significant for Bose 
 statistics ($q_{i}=m_i=0$ in Eq.(21)),  directions of 
arrows in such polygons are restricted.
  As shown in Fig.5, 6, and 7, there appear only special types of graphs, 
  in which every two vertices connected by a dotted line are separated by odd (not even) number of sides.  
  In combinatorial mathematics, the exact formula of $D(n)$ 
for such graphs was recently found \cite{nak}, and it has the following 
asymptotic form (Appendix.C) \cite{per}: 
\begin{equation}
	D(n)\rightarrow \frac{n!}{4n¥}, \quad as \quad n\rightarrow \infty  .¥
	\label{¥}
\end{equation}¥
Substituting Eq.(26) into Eq.(25) yields an inequality for $\Sigma D_n$, 
which will be discussed below.

\subsection{The instability of an attractive Bose gas in the normal phase}
Let us obtain an inequality for $Z_V(\mu)$ in low temperature. 
 Replacing $K_1^{\nu  _1} \cdots K_s^{\nu  _s}\cdots$ in Eq.(22) by $K_n$ 
 (as stated in (2) of Sec.3.A), and using Eq.(25) with Eq.(26) in Eq.(22), one obtains 
\begin{equation}
	 V\frac{1}{4n}K_n <\Xi _n< V\frac{1}{16\sqrt{3}n^2¥}\exp\left(\pi\sqrt{\frac{2n}{3¥}¥}\right)¥K_n¥¥¥.
\end{equation}¥
Using Eq.(23) in $K_n$ of Eq.(27) and applying it to Eq.(20), one obtains \cite{con} 
\begin{eqnarray}
	  \lefteqn{ Z_0(\mu)\prod_{p,l}\exp\left(V\sum_{n=1}^{\infty} 
                \frac{1}{4n}x(p,l)^n\right)¥¥¥  < Z_V(\mu) } \nonumber\\ 
        && < Z_0(\mu)\prod_{p,l}\exp\left(V\sum_{n=1}^{\infty}
	          \frac{\exp\left(\pi\sqrt{\frac{2n}{3¥}¥}\right)}{16\sqrt{3}n^2¥}x(p,l)^n \right).
\end{eqnarray}¥ 

 If the infinite series over $n$ in the exponent of the upper bound is 
 convergent for all $p$ and $l$, it 
 guarantees the finiteness of $Z_V(\mu)$, hence the stability of an attractive Bose gas. 
 Cauchy-Hadamard's theorem asserts that the radius of convergence $r_c$ of a power series  
$\sum_{n=1}^{\infty}a_nx^n$ is given by (see Appendix.D)
\begin{equation}
	\frac{1}{r_c¥}¥=\lim _{n\to \infty}|a_n|^{1/n}.
	\label{¥}
\end{equation}¥
 Applying this theorem to the upper bound
 [$a_n=\exp(\pi\sqrt{\frac{2n}{3¥}¥})/(16\sqrt{3}n^2)$] 
and the lower bound [$a_n=1/(4n)$] in Eq.(28), and using $\lim _{n\to 
\infty}n^{1/n}=1$, one obtains 1 as $r_c$'s of both bounds.
Hence, $Z_V(\mu)$ has the  radius of convergence $r_c=1$ as well. 
For each $p,l$,  the convergence condition $x(p,l) < r_c$ is given by
\begin{equation}
	      -U\frac{1}{\beta¥}¥¥
	 \frac{1}{(\epsilon_p-\mu)^2+\left(\frac{\pi l}{\beta¥}\right)^2}<1.
	\label{¥}
\end{equation}¥ 
The equation of state (1) and (2) implicitly determines the chemical 
potential of a gas as a function of its temperature  and pressure or density.
At high temperature, $\mu \ll 0$ is satisfied for the regions in 
$P-\rho$ space, and Eq.(30) is therefore
satisfied with respect to all $p$ and $l$. Hence, an attractive 
Bose gas ($U<0$) exists as the thermodynamically stable state.

With decreasing temperature, however, the negative $\mu$ 
gradually increases at a given $P$ or $\rho$. 
Changing $P,T$ or $\rho,T$ in $\mu$ of Eq.(30) makes possible the violation of the  
convergence condition. Among many $p$ and $l$, this condition is first 
violated at $p=l=0$ when  $\mu$ reaches a certain critical value $\mu _c$ as
\begin{equation}
 -\frac{U}{\beta \mu_c^2¥}=1,
	\label{¥}
\end{equation}¥
 where  $\mu _c$ satisfies
\begin{equation}
	\mu _c= -\sqrt {-Uk_BT_c¥} .
	\label{¥}
\end{equation}¥
 When $\mu$ reaches $\mu _c$, $Z_V(\mu)$ abruptly diverges without any 
 precursory behavior. Hence the specific volume $v$, determined by 
\begin{equation}
\frac{1}{v}=k_BT\lim_{V\to\infty}\frac{\partial}{\partial\mu}
ÊÊÊÊÊÊÊÊÊÊÊÊÊÊÊÊ \left(\frac{\ln Z_V(\mu)}{V}\right),
\end{equation}
discontinuously jumps to zero. This means that each 
particle cannot keep  large distances from other particles. Hence an 
attractive Bose gas undergoes a transition to the high-density state.

 {\it The condensation point $(P_c,\rho_c)$ in the isothermal curve is 
 determined by the equation of state and\/}
\begin{equation}
	\mu (\rho_c,T)=-\sqrt {-Uk_BT¥}
\end{equation}
{\it at a given temperature\/}\cite{nee}.

 For considering the influence of attractive force on the coherent 
many-body wave function, its critical size is of great interest. 
 Specifically, the critical size distribution $h(s)$ of elementary 
 polygons in Sec.2B is 
a significant quantity  \cite{gpl}. At high temperature ($\mu \ll  -k_BT$), $h(s)$ is a 
rapidly decreasing function of $s$. With decreasing temperature, $h(s)$ 
gradually changes to a weakly $s$-dependent distribution, finally 
reaching $h_c(s)$ at the condensation line
	\[ h_c(s) = \left\{ 
	  \begin{array}{ll}
	     \displaystyle{\exp\left(\frac{\mu_c}{k_BT_c¥}s¥\right)}&
	         \qquad p=0 \\[0.2cm]
	    \displaystyle{\frac{V}{\lambda^3¥}¥\frac{A_s}{s^{1.5}¥}¥\exp\left(\frac{\mu_c}{k_BT_c¥}s¥\right)}&      ¥
	         \qquad p\ne 0   ,
      \end{array}¥
     \right. \] 
 beyond which distribution an attractive Bose gas does 
not exist in the gas phase.  Hence, $h_c(s)$ is a {\it critical 
size-distribution \/} of the coherent many-body wave function in the gas phase.

The divergence of $Z_V(\mu)$ occurs first in bosons with zero momentum. This 
 means that {\it among many bosons with various momentums, the bosons with zero momentum
escape first from a gas, and make a liquid droplet at a certain point  of 
the gas\/}. Since this droplet is a high-density boson with $p=0$, it is 
likely that the growth of this droplet leads to a liquid in the BEC 
phase. As stated in Sec.1, in thermal equilibrium a Bose gas in the BEC phase undergoes no GLC, 
whereas {\it its GLC in the normal phase is likely to trigger the formation of 
the BEC condensate as a by-product\/}. The GLC of a normal helium 4 gas to a superfluid liquid at 
$P<0.05$ atm falls into this category. (One finds a parallel on this 
point in the BCS transition, in which the formation of loosely bound 
pairs of fermions is promptly followed by the formation of their BEC condensate.)

\subsection{Weak $U$ limit}
In contrast with the classical GLC, the weak-coupling limit ($U\simeq 
 0$) is one of the realistic situations of the {\it quantum GLC\/}.
 Even if $-U$ is very small, the condition (34) is satisfied for a 
 small $\mu$ at the vicinity of the BEC transition point in the normal phase. 
 In principle, an arbitrarily weak attractive interaction  creates the {\it 
 quantum GLC\/} at sufficiently low temperature.

 When we approximate $\mu$ in Eq.(34) with $\mu _0$ of an ideal Bose gas, we 
can make a rough estimation of $T_c$ and $\mu _c$ under the fixed density. In an ideal Bose gas, 
Eq.(2) is expanded as \cite {rob}
\begin{equation}
	\frac{\lambda ^3}{v¥}=g_{3/2}(e^{\beta\mu})\simeq 2.612-3.545\sqrt{|\beta\mu (T)|} ¥,
	\label{¥}
\end{equation}¥
where $g_{a}(x)=\sum_{n}x^n/n^{a}¥$.  Using Eq.(34) in $\mu (T)$ of the 
right-hand side of Eq.(35) at $T_c$, and dividing both sides with $\lambda 
_0^3/v=2.612$ (an ideal Bose gas), one obtains
  \begin{equation}
  	\left(\frac{\beta _c}{\beta _0¥}¥\right)^{1.5}
  	      +1.36\sqrt[4]{\frac{-U}{k_BT_0¥}¥}\left(\frac{\beta _c}{\beta _0¥}¥\right)^{0.25}-1=0.¥
  	\label{¥}
  \end{equation}¥  
 At $U\simeq 0$, one gets
\begin{equation}
  	T_c\simeq T_0 \left[1+0.90\sqrt[4]{\frac{-U}{k_BT_0¥}¥}\right]¥,
  	\label{¥}
\end{equation}¥ 
 and 
\begin{equation}
  	\mu_c\simeq -\sqrt{-Uk_BT_0¥}\left[1+0.45\sqrt[4]{\frac{-U}{k_BT_0¥}¥}\right]¥.
  	\label{¥}
\end{equation}¥ 
($T_0$ denotes the BEC transition temperature of an ideal Bose gas at the 
same $\rho$ \cite{eqr}.)
   
Figure.8 shows a schematic phase diagram. The dotted curve depicts the $P-T_0$ 
curve of  an ideal Bose gas, $P=(m/2\pi\hbar^2)^{1.5}g_{5/2}(1)(k_BT_0)^{2.5}$. 
  The solid curve illustrates  the $P-T_c$ curve (condensation line) 
  derived by applying  Eq.(37) to the $P-T_0$ curve. 
As a crude estimation of $T_{BEC}$ (traced back to London), we use $\lambda ^3/v =g_{3/2}(1)$ 
 with the experimental value of liquid density, and get 
$k_BT_{BEC}=2\pi \hbar^2 m^{-1}2.612^{-1/3}(N/V)^{1/3}$. In Fig.8,  
the  solid curve and a straight line $T=T_{BEC}$ intersect at the triple  
point. This schematic diagram is useful only for conceptual 
understanding, and comparing it with the phase diagram of the 
helium 4 needs elaborate calculations of some quantities, especially of the liquid state.

\section{Discussion}

\subsection{Comparison  with the classical GLC}
In the phase diagram of helium 4 like Fig.1, with increasing pressure, the 
nature of the GLC continuously changes from the quantum to the classical 
GLC \cite{cri}.  Let us compare the classical and the {\it quantum GLC\/}. 
 Since this paper considers only the attractive interaction for the 
 {\it quantum GLC\/}, we can make comparison only on the initial stage 
of the instability.  In the classical GLC, 
 Mayer's method considers the singularity  which must appear in the perturbation 
expansion of $Z_V(\mu)=\Sigma a_nx^n$.  
In general,  the relationship between the coefficient $a_n$ 
and the interaction potential is so complicated that it is a difficult 
problem whether the  $a_n$ derived from a given potential actually leads to 
such a singularity \cite{gro}.

(a) In a classical gas, since the GLC occurs at relatively high temperature, 
the inelastic scattering is important. Hence, one must perform an $(s-1)$-dimensional 
integral for obtaining the cluster integral. 
In the GLC of a dilute Bose gas,  however, since it occurs at very low temperature, 
 the diagram with a common momentum $p$ like Fig.5 becomes important. 
Hence, the integrals of a $2s$-sized polygon are factorized, and 
the bounds for $Z_V(\mu)$ become simple as in Eq.(28).

\begin{figure}
\includegraphics [scale=0.35]{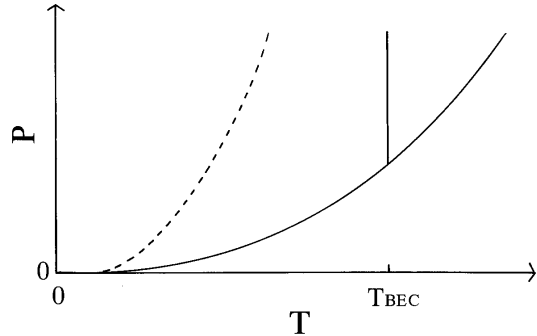}
\caption{\label{fig:epsart}
A schematic phase diagram. The dotted curve 
 is the $P-T_0$ curve of an ideal Bose gas at the BEC point. 
 The solid curve is the condensation line, that is, 
 the $P-T_c$ curve obtained by applying Eq.(37) to the $P-T_0$ curve. }
\end{figure}

(b) In a classical gas, one can prove the convergence or the divergence of 
$Z_V(\mu)$ only after summing up all momentum components.
In an attractive Bose gas, however, the instability is expected to occur at 
bosons with zero momentum. Hence, one can prove the divergence of 
$Z_V(\mu)$ by focusing on the $p=0$ component of  $\Xi _n$.  

(c) In the classical GLC, due to the absence of Bose statistics, there 
are far more $K_1^{\nu  _1} \cdots K_s^{\nu  _s}$-type clusters than 
$K_n$-type polygons. Hence, at each 
order of the perturbation, one cannot estimate $Z_V(\mu)$ only by the 
largest polygons in size. For the quantum GLC, however, one can 
estimate $Z_V(\mu)$  by focusing the largest polygons as in Sec.3.B.

These features (a) $\sim$ (c) will become useful for elucidating the 
essence of the {\it quantum GLC \/} and studying the classical-quantum crossover 
of the GLC in future studies.

\subsection{The quantum GLC in fermions}
 There is another type of {\it quantum GLC \/} occurring in 
fermions such as a helium 3 gas.  In the phase diagram of helium 3, the 
location of the normal-superfluid transition in the liquid phase is far 
apart from the condensation line between the liquid and the gas 
phase. (This is in contrast with helium 4 in Fig.1.) 
Hence, it is  uncertain to what extent the observed GLC is affected 
by Fermi statistics \cite{wel}.  In principle, however, the {\it quantum GLC\/}
in fermions has some different features from that in bosons. 
 For an ideal Fermi gas, due to one-particle excitation, the 
antisymmetrical coherent wave functions are not formed even at zero 
temperature, whereas it shows the Cooper instability with the attractive force.
Hence, it is appropriate to begin with the  pairing interaction when  
considering the  GLC.  In a Fermi gas, 
 a similar argument to that in Sec.2.C is valid  as well: 
When two Fermi particles ($p$ and $p'$) belonging to two different 
bubbles have the same momentum ($p=p'$), a new diagram containing 
a square must be included with a negative sign \cite{gau} \cite{lan}.
  Its $K_s$  has a similar form to Eq.(23) except for an 
odd integer $l$, and $\Xi _n$ of an attractive Fermi gas has a similar structure  
to Eq.(22) in appearance. But the argument in Sec.3.A is affected by Fermi statistics as follows. 

(a) Due to Fermi statistics, each $K_s$ in Eq.(22) has its own   
  momentum and frequency even at zero temperature, and 
 the number of the $K_1^{\nu  _1} \cdots K_s^{\nu  _s}$-type cluster 
 is therefore  larger than that of the $K_n$-type polygon in  $\Xi _n$. 
Hence,   at a given $n$,  dominant terms in the perturbation series 
 do not come from the $K_n$-type polygon, but from the 
various $K_1^{\nu  _1} \cdots K_s^{\nu  _s}$-type polygon clusters. 
 (In n=7 for example, the polygon cluster like Fig.4 is more important than 
the single largest polygon like Fig.5.) In this respect, the GLC in a 
Fermi gas has the same feature as the classical GLC [see (c) in Sec.4.A].

(b) For Fermi statistics, a negative sign appears in front of $U$ in $K_s$ 
(Eq.(21)), because of $a^{\dagger}_{p}a^{\dagger}_{-p}a_{-p'}a_{p'}=
- a^{\dagger}_{p}a_{-p'}a^{\dagger}_{-p}a_{p'}$ for $p \ne  p'$.
 Accordingly, the negative sign in front of $U$ ($<0$) in Eq.(23) disappears.  Hence,
   {\it  $\Xi _n$ of an attractive Fermi gas is an alternating 
 series. \/} The perturbation series of an attractive Fermi gas has a  
 completely different structure from that of an  attractive Bose gas.   
 That is, one must consider the cancellation of terms which are opposite in sign, 
 and cannot  approximate   $\Xi _n$ only by the terms with large 
$D_n$. One must carefully count all $K_1^{\nu  _1} \cdots K_s^{\nu  _s}$-type clusters
in Eq.(22). Gaudin \cite{gau} and Langer \cite{lan} gave an 
approximation of  $Z_V(\mu)$ along this line of thought, and led the BCS gap 
equation from it.   (Using their result, a rigorous proof was given recently that, as long as 
 attractive force is increased within the BCS scheme, the GLC is impossible \cite{koh97}.)
 Application of the inequality method to an attractive Fermi gas is a future problem.

\subsection{Future problems}
 Lastly, we point out some problems in future studies of the {\it quantum GLC \/}.

(a)  For Mayer's method of condensation, we know a long-standing 
problem in the interpretation of results.  Since this method focuses on 
the instability of a gas and contains no information on a liquid, there remains a question 
whether the singularity of the  power series in $Z_V(\mu)$ actually gives 
the condensation temperature $T_c$, 
or  corresponds to an end point of supersaturation $T_s$.
 This problem is related with technical difficulties in estimating the grand partition 
function.  In general, it is difficult to treat the highly inhomogeneous 
system so that it will reach true thermal equilibrium.  Poor approximations of  
$Z_V(\mu)$ often include the uniform-density restraint, thus 
preventing the system from reaching true thermal-equilibrium.  
The singularity of $Z_V(\mu)$ obtained by such an approximation 
corresponds to the end point of supersaturation. 
If a satisfactory approximation is made, it will lead to the condensation 
point in thermal equilibrium. 
Hence, the validity of Mayer's method depends on the type of phenomenon.
For bosons, we can have the following optimistic view:  
The uniformity of the system by $p=0$ particles is the characteristic property of bosons 
at low temperature. Hence the satisfactory approximation of $Z_V(\mu)$ is more probable 
 than that of more inhomogeneous systems.  A quantitative analysis of this point is a future problem. 
  In this paper, we used the term ``$T_c$ of the GLC'' as a generic name for the 
temperature at which a Bose gas shows the instability leading to the collapse.

(b) To discuss a liquid on a common basis with a gas, one must make a more realistic 
model by including the short-range repulsive interaction into 
Eq.(3). In this case, one must expect interference of the  attractive $U$ and the repulsive 
$U'$ in $\Xi _n$,  with a result that $\Xi _n$ becomes intermediate between the positive-term 
series and the alternating series.  An estimation of such a series  is 
a difficult problem.

(c) For the classical GLC, there appears long-range correlation at the 
vicinity of the condensation point, which comes from the general 
mechanism of a first-order phase transition.  For the {\it quantum GLC 
\/} in an attractive Bose gas, the long-range coherence comes from Bose statistics. This 
difference  may affect the nature of fluctuations in the vicinity of a transition point.  

(d)  The recent experiment of trapped atomic Bose gas  will
provide us with a new tool \cite{par}. 
Ultralow temperature is available and  direct control 
of the effective interparticle force becomes possible using magnetic 
field.  The system of interest from our viewpoint is not a BEC gas, 
but a normal gas just above the BEC transition point  \cite{dil}.
The trapped atomic Bose gas, however, has its own properties which may complicate 
the foregoing arguments, that is, metastability.  The confinement of atoms by a harmonic potential 
gives rise to the spatial inhomogeneity in the system. Many bosons gather at the center of a trap.  
The resulting increased density enhances the zero-point motion of particles, which stabilizes the gas 
state.  For a trapped atomic Bose gas with attractive interaction, one finds a 
metastable BEC state   at low temperature and below critical density. Up 
to now, the collapse of such a metastable BEC gas was detected 
\cite{bra} \cite{cor}, but the GLC in the normal phase has not been 
observed. Further, there is a possibility that a state after the transition is not a liquid but a 
solid or molecules. If it is true, we must consider this system from a 
different viewpoint.

\section*{Acknowledgment}
The author thanks Professor O.Nakamura for combinatorial mathematics.

%\newpage 
\appendix

\section{Comments on Ref.[8]}
 As a starting point to $Z_V(\mu)$ of an attractive Bose gas,  
 Ref.\cite{koh01} began with an approximate formula [Eq.(9) in Ref.\cite {koh01}], 
 which carefully deals with the $K_1^{\nu  _1} \cdots K_s^{\nu  _s}$-type 
 cluster like Fig4.  As confirmed in Sec.3, however, the peculiar nature of a
Bose gas appears in the $n\rightarrow \infty$ limit. 
In this limit, Eq.(9) and all subsequent
integrals (with respect to $t$) in Ref.\cite {koh01} turn out to diverge. 
 The finial form of $Z_V(\mu)$ (Eq.(21) in Ref.\cite {koh01}) approaches $\infty -\infty$, 
 which loses the mathematical meaning.
 Hence, it is questionable to identify it as a Yang-Lee zero. 
(One way to avoid this divergence is to perform all integrals in Ref.\cite 
{koh01} over $[0,M]$ with a finite $M$ instead of $[0,\infty]$. 
But such a trick is unnatural in a Bose gas.)
In this paper, we prove the singularity of $Z_V(\mu)$ by showing its 
divergence using the inequality; this gives a sound basis to the 
instability condition (Eq.(34)), which was first obtained in Ref.\cite {koh01}.

\section{polygons in $Z_0(\mu)$}

 $f_s$ in Eq.(9) is estimated as follows \cite{fey}. If $x_{s1}^2$ 
[the last term in the exponent of Eq.(9)] is replaced by $x_{s0}^2$,  $f_s$  
becomes a function of $x_1-x_0(\equiv x_{10})$ as $f_s(x_{10})$. 
Equation (9) shows that $f_s(x_{10})$ is made by $(s-1)$ times of the 
convolution of $\exp (-mx^2/2\beta\hbar^2)$. Hence, 
 using its three-dimensional Fourier transform 
 $\Gamma(p) = \lambda'^3\exp (-\lambda'^2p^2/2)$, where 
 $\lambda'=\lambda/\sqrt{2\pi}$, $f_s(x_{10})$ is expressed as 
 \begin{equation}
 f_s(x_{10})= \frac{V}{(\sqrt{2\pi})^3¥}\int (\sqrt{2\pi})^{3(s-1)}\Gamma(p)^s
                 e^{-ipx_{10}}4\pi p^2dp¥,
\label{}
\end{equation}
where $V$ corresponds to the translation of the elementary polygons.
Since $f_s=f_s(x_{10}=0)$, one has 
 \begin{equation}
 f_s= V\int (\sqrt{2\pi})^{3s}\Gamma(p)^s\frac{4\pi p^2dp}{(2\pi)^3¥}¥.
\label{}
\end{equation}
After extracting $\Gamma(p=0)$ from the integrand of Eq.(B2) and 
performing the integral, one obtains
\begin{equation}
 f_s= \left(\lambda ^{3s}+A_s\lambda ^{3(s-1)}\frac{V}{s^{3/2}¥}¥¥\right)¥.
\label{}
\end{equation}

\section{asymptotic formula of $D$}
The asymptotic form of $D(n)$ (Eq.(26)) is derived as follows \cite{nak}.
  We think of regular $2n$-polygons with $n$ interaction 
lines, in which $n$ lines connect $n$ pairs of vertices separated by an
odd number of sides. We denote its number by $D(n)$, and classify it by a 
 transformation property such as  rotation and reflection: 
 A polygon is invariant under rotations through angle $\pi l/n$ 
($l=0,1,\ldots , 2n-1$). Further,  a $2n$-polygon is symmetrical with 
respect to the $2n$ axis. 
 Hence the symmetry group is composed of $4n$ elements $C_0,C_1, \ldots 
 ,C_{4n-1}$. We denote by $\phi (C_i)$ the number of graphs 
which are invariant under transformation $C_i$.

 Birnside's theorem \cite{liu} states  $D(n)=\Sigma \phi (C_i)/4n$.
 The most important $\phi (C_i)$  is  the number of polygons invariant under an 
 identical transformation $\phi (C_0)=n!$ (the number of ways 
 of connecting $n$ pairs of vertices).  Among many $\phi (C_i)$'s, the 
 $\phi (C_0)$ becomes dominant as  $n\rightarrow\infty$.  Hence,  
taking only $\phi (C_0)$ in $\Sigma \phi (C_i)$, one obtains an asymptotic 
form  $D(n)\rightarrow n!/4n$ at $n\rightarrow\infty$. 

For $D(n)$ with a finite $n$, one must classify $2n$-polygons 
with  interaction lines by symmetry, and include other $\phi (C_i)$'s  
in $\Sigma \phi (C_i)$.
 The result is as follows \cite{nak}.
\newtheorem{th1}{Theorem}

\begin{th1}[Nakamura] The number of the nonequivalent 1-regular odd 
spanning  subgraphs of the complete graph of order 2n by the Dihedral 
group $D_{2n}$ is given by

\[   D(n)=\frac{n!}{4n¥}
            +\frac{1}{4n¥}\sum_{i=1}^{2n-1¥}R^{2n}_i
           +\frac{1}{4}(S_0+S_1)¥¥ , \]
           
where

  \[R_i^{2n}=\left\{
     \begin{array}{ll} 
         \displaystyle {0}, & ¥¥
             \quad d=odd,  n=2m\\  \\ [0,2cm]
         \displaystyle {\sum_{d=2s+t}¥\frac{d!}{2^ss!t!¥}¥ 
                         \left(\frac{n}{d¥}¥\right)^{s} }, &
                         \quad d=odd,  n=2m+1\\  \\ [0,2cm]
         \\ 
         \displaystyle {\left(\frac{2n}{d¥}¥\right)^{d/2}\left(\frac{d}{2¥}¥\right)!}, & ¥¥
             \quad d=even, 
     \end {array}  \right. \] 

(d is the greatest common divisor of 2n and i, and s and t are positive).
\[ S_0=\left\{
     \begin{array}{ll} 
         \displaystyle {0}, & ¥¥
             \quad n=2m \\ [0,2cm]
         \displaystyle {2^{(n-1)/2}\left(\frac{n-1}{2¥}¥\right)!¥¥}, & ¥¥
             \quad   n=2m+1, 
      \end {array}  \right. \]  

\[ S_1=\sum_{n=2s+t}¥\frac{n!}{2^ss!t!¥}¥ 
                          \quad .  \]

\end{th1}¥

\section{Cauchy-Hadamard's theorem}
We begin with a general series  $\sum_{n=1}^{\infty¥}c_n$. 
If one can find $k$ and $m$ such that $\sqrt[n]{c_n}\leq k<1$  when $n\geq m$, one gets for a large $n$
\begin{equation}
	S_n=c_1+\cdots +c_n<k+\cdots +k^n=\frac{k(1-k^n)}{1-k¥}¥.
	\label{¥}
\end{equation}¥
Since $0<k<1$, $\sum_{n=1}^{\infty¥}c_n$ converges. 

The above result is used in $\sum_{n=1}^{\infty¥}a_nx^n$ with $|a_nx^n|$ 
identified as $c_n$. By defining $\lim_{n\rightarrow\infty}|a_n|^{1/n}\equiv l$, 
one obtains $\lim_{n\rightarrow\infty}|a_nx^n|^{1/n}=l|x|$, that is,
 $\lim_{n\rightarrow\infty}\sqrt[n]{c_n}=l|x|$. 
 
This means that, if $l|x|<1$, Eq.(D1) is valid, hence $\sum_{n=1}^{\infty¥}c_n$  
converges. Contrarily, if $l|x|>1$, it diverges. 
The radius of convergence $r_c$ of $\sum_{n=1}^{\infty¥}a_nx^n$ is given by
$l|r_c|=1$, that is,
   \begin{equation}
   		\frac{1}{r_c¥}¥=\lim _{n\to \infty}|a_n|^{1/n}.
   	\label{¥}
   \end{equation}¥

¥

%\end{references}



\begin{thebibliography}{99}
\bibitem{wel}  At 1 atm, however, the GLC of a helium 4 gas to a normal liquid is 
         regarded as the classical GLC ($T_c=4.215 K$). 
         At 1 atm, the interaction between helium  atoms is not so weak that  
         the GLC occurs at relatively high temperature, thus washing out 
         the feature of quantum statistics.
        (A helium 3 gas at 1 atm undergoes the GLC at $T_c=3.191 K$. 
        The small difference of $T_c$ between $^4$He and $^3$He suggests 
        that quantum statistics plays a minor role.)
\bibitem{coh} Note that the coherence size is a different concept 
    from the coherence length (healing length).
\bibitem{rep} In a repulsive Bose gas ($U>0$), the sign of $K_s$ alternates 
    as $s$ increases. Hence Eq.(22) becomes an  alternating series, making 
     the estimation of perturbation series difficult.
\bibitem{per} After one of $H_{it}(\beta)$ is fixed in
     $<T H_{it}(\beta _1)\cdots H_{it}(\beta _n)>$, making all 
    possible permutations of $n-1$ operators $H_{it}(\beta)$ yields 
    $(n-1)!$ (circular permutation), which 
    agrees with the order of $D(n)$  at $ n\rightarrow \infty$ in Eq.(26).
\bibitem{con} Both bounds have a form of $\exp (aV)$, which $Z_V(\mu)$ must 
     approach at $V\rightarrow \infty$ as generally proved in Ref.\cite{yan}.
\bibitem{nee} Since the BEC phase is excluded from a list of possible 
      states of an attractive gas in thermal equilibrium, there is no need to compare 
      Eq.(34) with the BEC condition.
\bibitem{gpl} In addition to $h(s)$, the size distribution of 
      interaction polygons is another important quantity, 
      and we denote it by $g_{p,l}(s)$.  It appears in the pressure of a gas as 
      $P=(k_BT/V)\left[\sum_{s}h(s)/s +\ln\left(\sum_{p,l}\sum_{s}g_{p,l}(s)\right)\right]$. 
     In contrast with $h(s)$, many $g_{p,l}(s)$'s with different $p,l,s$  
      are connected to each other in the logarithm.  Hence, $g_{p,l}(s)$ is not as simple as $h(s)$.
\bibitem{eqr} Equation (32) in Ref.\cite {koh01} must be replaced by Eq.(38) in this paper.
\bibitem{cri} The critical point (the end point of 
       the condensation line) is an attribute of the classical GLC.
\bibitem{dil} In a low-density gas, the different wave functions overlap only 
         slightly.  At the vicinity of the BEC phase, however, the emergence  
         of the large coherent many-body wave function completely changes the 
         situation.   In fact,  a trapped atomic BEC gas cooled by optical techniques is 
         $10^3\sim 10^6$ times as dilute  as the measurable low-pressure limit of
          helium 4 gas cooled by the conventional method.
\end{thebibliography}
\end{document}